\documentclass{article}

\usepackage[square,numbers]{natbib}
\usepackage[final]{neurips_2021}
\usepackage{caption,tabularx,booktabs}
\usepackage[utf8]{inputenc} 
\usepackage[T1]{fontenc}    
\usepackage{hyperref}       
\usepackage{url}            
\usepackage{booktabs}       
\usepackage{amsfonts}       
\usepackage{nicefrac}       
\usepackage{microtype}      
\usepackage{xcolor}         
\usepackage{graphicx}       
\usepackage{float}          
\usepackage{subfigure}      
\usepackage{mhchem}         
\usepackage{algorithm}      
\usepackage{algorithmic}    
\usepackage{multirow}       

\title{Interdisciplinary Discovery of Nanomaterials Based on Convolutional Neural Networks}

\author{
  Tong Xie\\
  University of New South Wales\\
  \texttt{tong.xie@unsw.edu.au}
  \And
  Yuwei Wan\\
  City University of Hong Kong\\
  \texttt{yuweiwan2-c@my.cityu.edu.hk}
  \And
  Weijian Li\\
  Northwestern University\\
  \texttt{weijianli2021@u.northwestern.edu}
  \And
  Qingyuan Linghu, Shaozhou Wang, Yalun Cai\\
  University of New South Wales\\
  \And
  Clara Grazian\\
  University of Sydeny \\
  \texttt{clara.grazian@sydney.edu.au}
  \And
  Han Liu\\
  Northwestern University\\
  \texttt{hanliu@northwestern.edu}
  \AND
  Chunyu Kit\\
  City University of Hong Kong\\
  \texttt{ctckit@cityu.edu.hk}
  \And
  Bram Hoex\\
  University of New South Wales\\
  \texttt{b.hoex@unsw.edu.au}
}

\begin{document}

\maketitle

\begin{abstract}
  The material science literature contains up-to-date and comprehensive scientific knowledge of materials. However, their content is unstructured and diverse, resulting in a significant gap in providing sufficient information for material design and synthesis. To this end, we used natural language processing (NLP) and computer vision (CV) techniques based on convolutional neural networks (CNN) to discover valuable experimental-based information about nanomaterials and synthesis methods in energy-material-related publications. Our first system, TextMaster, extracts opinions from texts and classifies them into challenges and opportunities, achieving 94\% and 92\% accuracy, respectively. Our second system, GraphMaster, realizes data extraction of tables and figures from publications with 98.3\% classification accuracy and 4.3\% data extraction mean square error. Our results show that these systems could assess the suitability of materials for a certain application by evaluation of synthesis insights and case analysis with detailed references. This work offers a fresh perspective on mining knowledge from scientific literature, providing a wide swatch to accelerate nanomaterial research through CNN.
\end{abstract}

\section{Introduction}

The scientific literature is meant to communicate recent advances and keep researchers informed so they can continue to make breakthroughs. Since innovative materials have been regarded as one of the promising roles leading the fourth industrial revolution, the material science field has witnessed a drastic volume increase in publications. Therefore, comprehensive selection and reading of the material science scientific papers by hand are not realistic. A potential solution to this issue is to automatize the task using artificial intelligence (AI) technologies. However, the information in scientific literature is multimodal (text, figures, tables, etc.), which makes utilizing latent knowledge in publications even more challenging.. 

In this paper, we resort to CNN to discover and summarize knowledge related to nanomaterials from scientific literature, with a novel perspective of combining the NLP and CV techniques to assimilate and integrate the multimodal literature knowledge. Firstly, we applied the classic NLP task opinion mining to the body content of energy-related publications. The system includes four modules: (i) text preparation,
(ii) opinion extraction, (iii) opinion classification and (iv) opinion mining for information analysis. This system can assess the suitability of nanomaterials for an application from various aspects, such as synthesis performance and natural properties. Secondly, we develop a system to assist in compiling a systematic and highly refined performance review of b-Si solar cells. The system includes (i) content retrieval, (ii) graph classification, (iii) data extraction. NLP techniques are applied to help screen out irrelevant articles in module (i). With modules (ii) using a CNN-based neural network, ResNet \cite{russakovsky2015imagenet, jung2017chartsense}, identifying the different elements, i.e., texts, tables, figures and images, then different NLP and CV models are further applied to extract corresponding data in module (iii). This system indicates that hierarchical textures and inverted-pyramidal black silicon (b-Si) nano-textures and micro-textures are promising as next-generation texturing techniques for Si solar cell applications.

\section{Related Works}
We identify related works for this paper in the area of NLP and CV methods. NLP methods are categorized into two domains, namely: NLP methods related to scientific discovery and opinion mining techniques. CV methods are mainly related to image recognition.

\paragraph{NLP methods related to scientific discovery}
NLP researchers have implemented different tasks on scientific text, such as pre-trained language model for scientific text \cite{beltagy2019scibert}, text classification \cite{baker2016automatic}, text summarization \cite{wang2021systematic}, named entity recognition (NER) \cite{perera2020named} and knowledge graph \cite{wang2021text}. In materials science, significant effort has been devoted to extracting material mentions \cite{swain2016chemdataextractor, weston2019named} or synthesis recipes \cite{kim2017virtual, huo2019semi, jensen2019machine}. These material science applications are based on NER, linking material names to their co-occurring entities, which help answer large-scale questions, such as "What are the most frequent methods for \ce{TiO2} synthesis?" Recent attempts to predict suitable materials for certain applications have been made by word embeddings \cite{tshitoyan2019unsupervised} or graph-based machine learning \cite{hatakeyama2020integrating}, which efficiently encodes materials science data into high-dimensional numeric representations.  

\paragraph{Opinion mining techniques}
Opinion mining is an NLP task that analyzes people’s opinions, sentiments, evaluations, attitudes, and emotions from written language \cite{liu2012sentiment}. It has widespread applications, such as categorizing product reviews to support consumers’ decisions \cite{haque2018sentiment}, analyzing social media texts to predict the stock market \cite{pagolu2016sentiment} or election results \cite{wang2012system}, and fostering effective interactions in smart cities \cite{ahmed2016sentiment}. This technique has been applied to scientific literature but mostly focuses on article reviews \cite{keith2019sentiment} or citations \cite{yousif2019survey} instead of the article’s body content. We creatively applied sentence-level opinion mining to the body content of energy material-related publications. Existing sentence-level opinion mining approaches can be divided into lexicon-based and corpus-based approaches \cite{mehta2018attention}. Lexicon-based approaches are usually unsupervised, based on the measurement of word sentiment orientation using WordNet or machine learning models \cite{hatzivassiloglou1997predicting, kamps2004using}. Corpus-based approaches are usually supervised, and dominated by feature-based machine learning methods and deep learning methods. In comparison, deep learning methods show better performance on most of the benchmark datasets. 

\paragraph{CV methods related to scientific graph classification}
The first graph type classification work was introduced by Zhou and Tan\cite{zhou2001learning} with a modified probabilistic through transform methods. They achieved 84\% accuracy of graph type classification in 1190 total images. After that, graph classification research began to attract attention. Scientists have applied different methods for graph classification problems, such as hidden Markov models, SVMs and CNNs. We summarize some of the existing graph classifications works in Table 1.

\paragraph{Techniques of data extraction from chart}
Previous research on data extraction from charts can be divided into two types, interactive and full automated approaches. Chartsense \cite{jung2017chartsense} was a semi-automated data extraction tool that provides the user with an interface to mark graph elements such as the x-axis and the origin of coordinates. However, interactive extraction was time-consuming and unsuitable for large data collection. Automated graph data extraction such as Revision \cite{savva2011revision}, Chartreader \cite{perera2020named}, and Chartdecoder \cite{dai2018chart} used different Optical character recognition (OCR) engines and CNNs to realize Bar Chart data recovery. However, for material science publications, a lot of important material information was presented with line charts, such as spectroscopy spectra and wavelength-related graphs. In this work, we demonstrated how to use CNN to fully automatically extract the data from the material science line charts.

\begin{table}[H]
    \centering
    \begin{tabular}{ |c|c|c|c|c| } 
    \hline
    Work & Methods & Type of Labels & Dataset & Performance \\
    \hline \cite{zhou2001learning} & Customized Algorithm & 3 & 1190 & 84\% \\ 
    \hline \cite{savva2011revision} & SVM & 10 & 2601 & 75.00\% - 86.00\% \\
    \hline \cite{dai2018chart} & CNN & 5 & 11174 & 99\% \\ 
    \hline \cite{jobin2019docfigure} & CNN & 28 & 33000 & 88.96\% - 92.90\% \\
    \hline
    \end{tabular}
    \caption{Summary of the related graph type classification work}
    \label{tab:my_label}
\end{table}

\section{Methods}
\label{gen_inst}

Here we introduce NLP and CV methods that were used to analyze energy-related publications. Our python code (includes TextMaster, GraphMaster and SciCrawler) and data is available at three repositories in https://github.com/EnergyMasterAI. 

\subsection{TextMaster}
Schematic diagram of TextMaster system is shown in Figure 1.
\begin{figure}[H]
  \centering
  \includegraphics[width=0.8\textwidth]{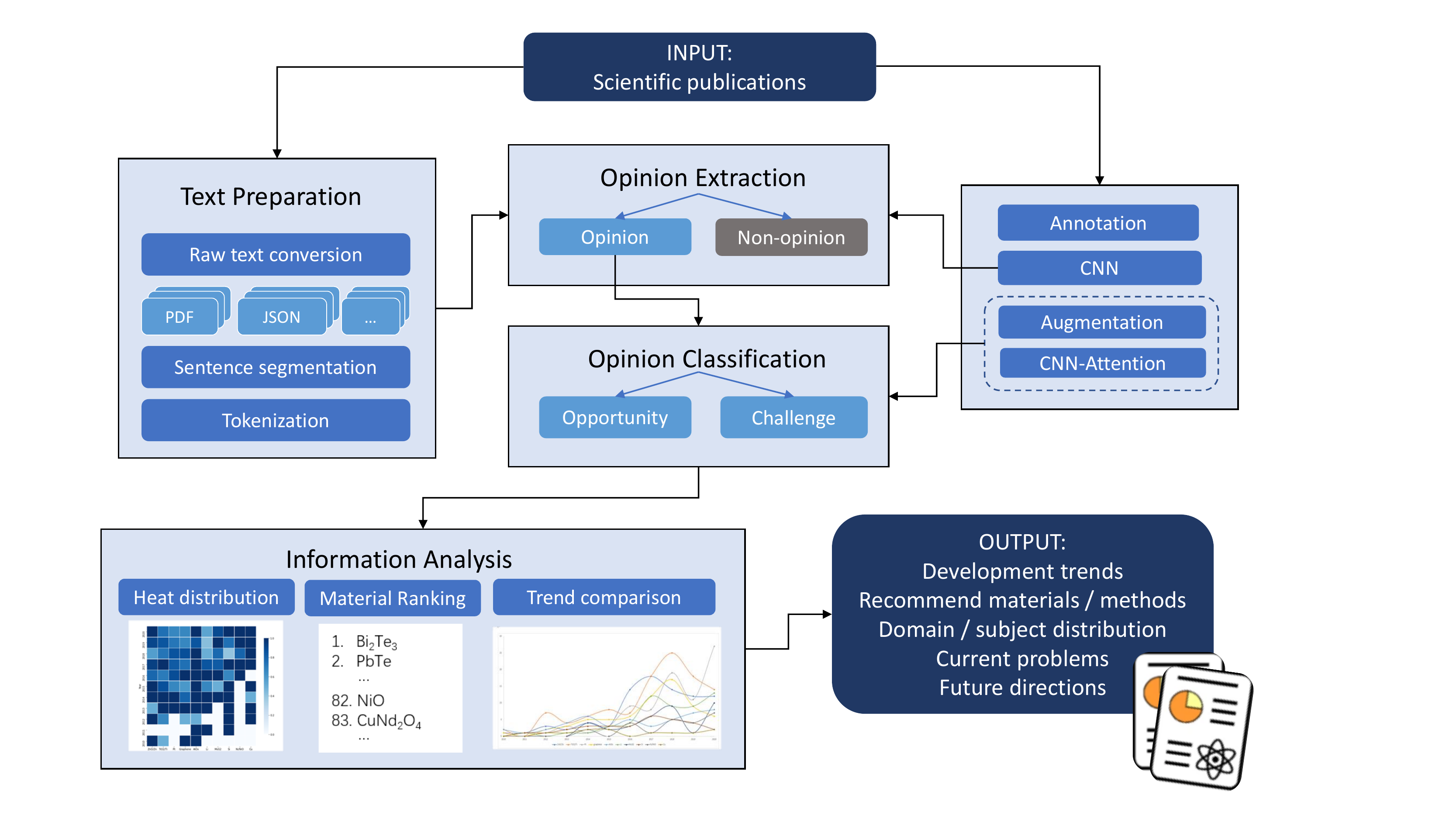}
  \caption{Schematic diagram of TextMaster system pipeline}
\end{figure}

\subsubsection{Text preparation}
We used the Web of Science client to collect DOI lists for various nanomaterial scientific articles. Then we customized a web-scraper, SciCrawler, to automatically download a broad selection of nano materials-related papers published after 2000 from publishers' API or websites in PDF format. The publications on a certain topic were collected by searching keywords from databases such as Web of Science. We prepared a mixed-topic dataset formed by three datasets with end-of-life management (EoL), perovskite solar cells, and atomic layer deposition (ALD) themes, covering management, material and synthesis. The EoL dataset was formed by 77 full papers. Perovskite and ALD datasets were formed by 34,752 abstracts and 21,276 abstracts respectively, containing all meta fields provided by Web of Science. Plain texts were extracted, segmented into sentences, and tokenized. The sentences in EoL dataset were annotated as "opinion" and "non-opinion", and "opinion" sentences were further annotated as "opportunities" and "challenges".

\subsubsection{Opinion extraction}
The processed sentences were classified as opinions and non-opinions in this module. In the pilot study, we compared the performance of some typical lexicon-based methods and deep learning models on the EoL data set (experiment details see Appendix A). It can be seen that the CNN model exceeds other methods on distinguishing opinions from non-opinions. Due to the imbalance of opinion and non-opinion categories, we also used the Synthetic Minority Oversampling Technique (SMOTE) \cite{chawla2002smote} to generate data for opinion category. To increase data size, the CNN-SMOTE (CNN with SMOTE technique) trained on the EoL dataset was used as raw models to predict the pseudo labels of sentences in abstracts of ALD- and perovskite-related publications. Each sentence was predicted as opinion or non-opinion. Then, annotators were invited to proofread the predictions. We trained the final CNN model on about 22 thousand samples, achieving an accuracy of 94\%.

\subsubsection{Opinion classification}
The input opinions were classified as opportunities and challenges in this module. Compared with CNN, a CNN model with attention mechanism  \cite{vaswani2017attention} had a better performance in the pilot study (see Appendix A). Similar to the last module, we used a raw model to increase data size and trained the final CNN-Attention model on about 9 thousand mixed-topic samples. Due to the imbalance of opportunity and challenge categories, we also used the SMOTE to generate data for the challenge category, which improved the performance of the CNN-Attention model from 89\% to 92\%.

\subsubsection{Information analysis}
With extracted and classified opinions, this module used embedded material names or elements for information analysis. For each opportunity or challenge sentence, ChemDataExtractor \cite{swain2016chemdataextractor} was used to identify the material name in it. Based on the sentiments of opinions and year of publication, we can calculate each material’s percentage of opportunities in opinions each year. This method quantified the sentiment towards elements as sentiment scores across time. Using opinions, we implemented a case study of ALD elements and a prediction of thermoelectric materials (see the Results and Discussion Section).

\subsection{GraphMaster}
Figure 2 presents the graph data recovery pipeline, which mainly consists of three components, i) content retrieval ii) graph classification and iii) data extraction. 
\begin{figure}[H]
  \centering
  \includegraphics[width=0.9\textwidth]{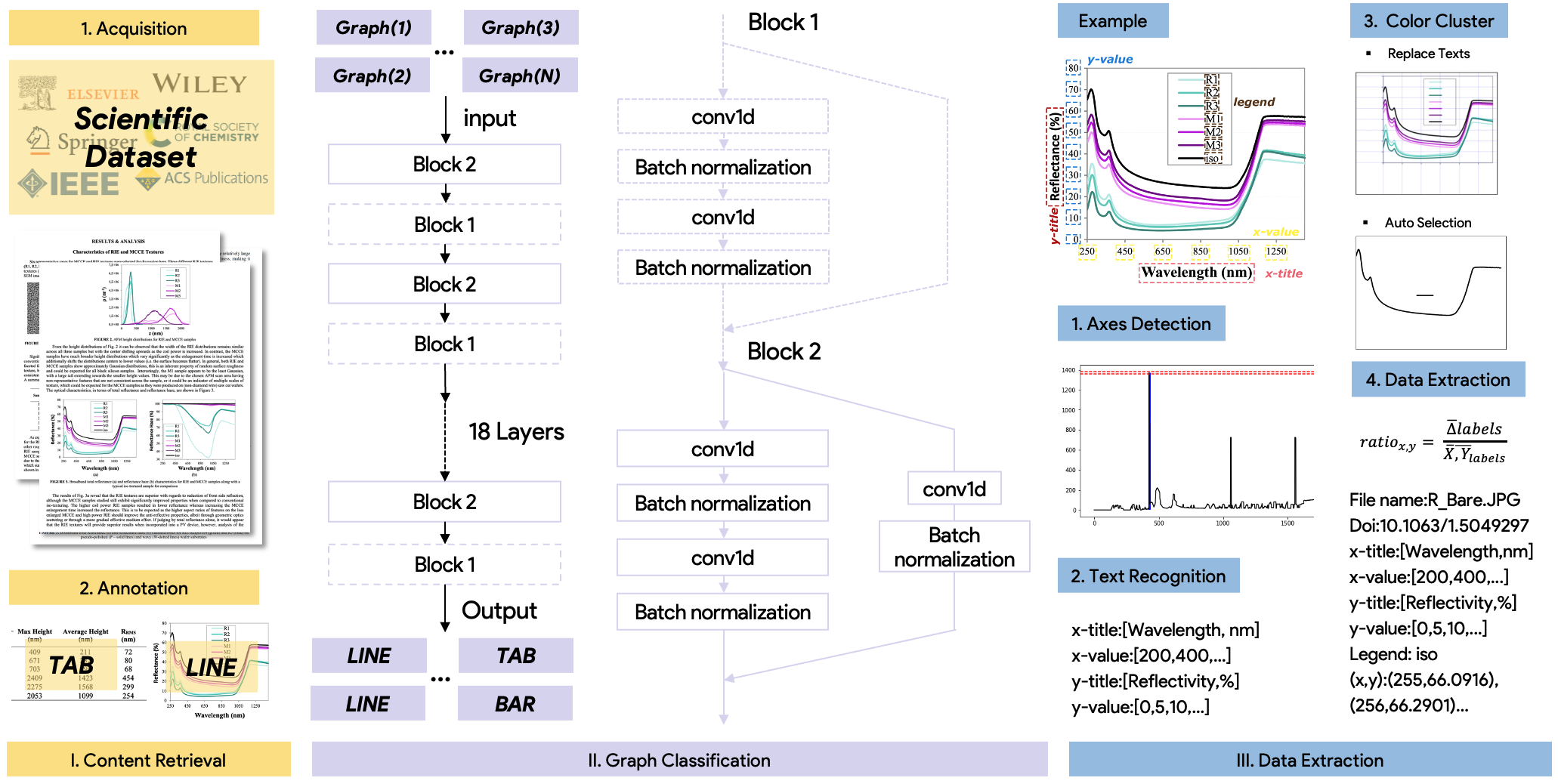}
  \caption{Schematic diagram of GraphMaster system pipeline for line chart data extraction}
\end{figure}
In this paper, we only proposed the data extraction solution for the line graph, since most of the material's characterization figures are line graphs including Raman spectroscopy, X-ray diffraction, external quantum efficiency graphs, etc. There are some existing solutions for data recovery of other graphs, including bar charts and tables
\cite{dai2018chart,swain2016chemdataextractor,mavracic2021chemdataextractor}. 

\subsubsection{Content Retrival}
The paper collection process is similar to the TextMaster (See 3.1.1). PDFFigure 2 \cite{mouchere2016advancing,mahdavi2019icdar} , an existing algorithm, was used to extract figures from scientific papers. In material science publications, one figure commonly includes several sub-figures. However, it is hard for PDFFigure 2.0 to capture each subfigure. We decomposed the figure into subfigures through valid axis-aligned splitting recognition\cite{siegel2016figureseer}. Finally, we manually assigned them category labels, including line charts, bar charts, and tables. The figure and metadata like the graphs, journal, and title were stored in the Neo4j dataset for convenient access (https://neo4j.com/). After the selection of samples related to the analyzed materials, we used 6,000 figures from 3,600 scientific papers. 

\subsubsection{Graph Classification}
We used a 20-layer ResNet \cite{he2016deep} with 9 bottleneck blocks to do binary classification on the collected graphs. The purpose of the classification was to extract the specific type of charts that were ready to be fed to the information detection step in the pipeline to extract the data of the plot. In this paper, the extraction step focused on the line charts. However, we tried to make every module as general and extensible as possible. Thus, the classification module is not only able to classify line charts, but also other three types of charts and images that were prevalent in academic literature. The four types were line charts, images of tables, bar charts and images of pseudo code of algorithms. The model was trained on about 20 thousand charts and images with labels using the cross entropy loss. We achieved very high accuracy on the classification task in all four categories, 97.9\% accuracy on line charts, 97.33\% accuracy on table images, 99.19\% on bar charts and 98.78\% on algorithm images. All the accuracies were obtained on a set-aside test dataset. This high accuracy on the test dataset suggests that our graph classification module has a very strong generalization performance. This solidly indicates that it would not only perform well on our own dataset but will also perform well on any unseen dataset of the same types, which means that our model would be practical and useful in real-world using scenarios. 

\subsubsection{Data Extraction}
\paragraph{Axes Detection}
Inspired by chartreader \cite{rane2021chartreader}, we convert the line charts to the grey-scale image to identify the location of the x-axis and y-axis, by replacing all pixels with luminance greater than 200 with value '1' (black) and all other pixels with value '0' (white). We experimented with different values for grey-scale thresholds and 200 was found to be best for grey-scale image conversion. For x-axis identification, we scanned the matrix horizontally the recorded the black pixels' continuity within surrounding rows. Then we selected all the continuous black pixels with less than ten white pixels in the row. Finally, we chose the column with the lowest fluctuation by the highest continuously of the adjacent rows. The reason was that that axis commonly consisted of 2-3 pixels in publication for clear reading. With a similar process, we can also detect the y-axis column. 

\paragraph{Text Detection}
AWS-Rekognition DetectText API was used to detect text within an image. DetectText also provided rectangular bounding boxes of the detected text. We selected the detected text with the highest confidence score. A high threshold of the confidence score improved precision. However, it would also reduce the recall rate. Therefore, in order to mitigate recall rate reduction, we added a dual layer for the second pass to release high-quality recognition. To this end, we set the confidence score threshold to 82 empirically. The bonding boxed centre coordinates and height and width were stored in a matrix, $\{t_x,t_y,t_w,t_h\}$, and they were replaced with white colour.

\paragraph{Text Extraction}
A line chart contains a lot of different texts, such as figure titles, axes titles, labels, and legends. The five text types that we classified were listed below:

\emph{X-axis value} (yellow): x-axis text is the bounding boxes detected below the x-axis, and their centroids are parallel to the x-axis. 

\emph{X-axis title} (pink): x-axis text is the bounding boxes detected below the x-axis title, which is near the bottom of the figure in the relative coordination.

\emph{Y-axis value} (blue):  y-axis text is the bounding boxes detected left to the y-axis, and their centroids are parallel to the y-axis. 

\emph{Y-axis title} (red):  y-axis text is the bounding boxes detected left to the y-axis, and their centroids are parrel to the y-axis.

\emph{Legend} (brown): legends are the bounding boxes detected in other locations.

\paragraph{Colour Cluster}

Since the line charts usually contained many lines, the authors used different features to distinguish, such as colours and dash lines. We applied the DBSCAN \cite{ester1996density} to classify different figures based on the RGB value of each colour. We examined the classification results on a set of randomly selected groups, and 16 groups had the best colour classification results. We selected the figures with the most pixels, and the threshold depended on the number of legends bounding boxes detected. (See Appendix C for an example diagram detailed process)

\paragraph{Data Recovery}
Based on the clustering results, we recovered data from each line consecutively. Specifically, for each line, we scanned from left to right horizontally. At each iteration (at each x value), we examined the vertical column of pixels. There was always one coloured pixel that belongs to the line while all other pixels belong to the white background. Besides, we can easily locate the position of that coloured pixel in that column. By calculating the ratio between the coloured pixel's height and the total height of the column, we can recover the y value for this specific x value. By repeating the same process for all x values in the scan and for each line in the plot, we can recover all the x-y value pairs of the plot. (See Appendix D for an example detailed process)

\section{Results and Discussion}
\label{headings}
Here we display the results of two systems on different publication datasets and discuss how they can render the latent knowledge and accelerate the energy material research.
\subsection{CNN assisted trends analysis and insights discovery for nanomaterial synthesis}
In this paper, we showed the preliminary findings of ALD, which was a popular nano-film fabrication technique. The ALD dataset had over 2,100 opinions with material names, in which nearly 70\% of their sentiments are opportunities. To avoid data fragmentation, we clustered opinions to corresponding periodic table elements of materials. We visualized scientific categories of the top 10 elements used in ALD in Figure 3. It can be summarized that popular chemical elements used in ALD are widely employed in material science, chemistry, nanoscience and physics, and to a lesser extent, energy and fuels. In addition, categories like crystallography and pharmacy are unique to the ALD of specific elements, namely, titanium (Ti). The gap between disciplines may hinder the extraction and utilization of shared knowledge about materials. This figure implies the great potential to transfer knowledge in opinions from one field to another.

We visualized the frequency of popular elements used by opinions from 2014 to 2022 in Figure 4(a). To quantify the sentiment towards elements, we calculated each element’s percentage of opportunities in opinions each year based on the sentiments and year of publication. As shown in Figure 4(b), each cell represents the average sentiment value of the element fabricated with ALD that year. We observed that there is often a drop in frequency after the year when the element experienced a rather negative sentiment and vice versa. For example, sentiment scores of Zn (dark blue), Ti (orange), Mo (pink), and Si (grey) had low levels in 2018, and their frequency fell in the following years. In contrast, the sentiment score of Mo (pink) in 2016, Ni (yellow) in 2017 and Al (purple) in 2019 were all high levels, and their frequency went up after experiencing positive sentiments. Although this relationship cannot be guaranteed, it helps us to predict the future trends of elements: Al, Si and Ni go rather cold; Pt, C and Cu emerge more; Zn, Ti, Li and Mo have relative stable attention. We also visualized opinion mining results of elements or materials in other domains such as perovskite and thermoelectric (see Appendix B).
\begin{figure}[H]
  \centering
  \includegraphics[width=0.8\textwidth]{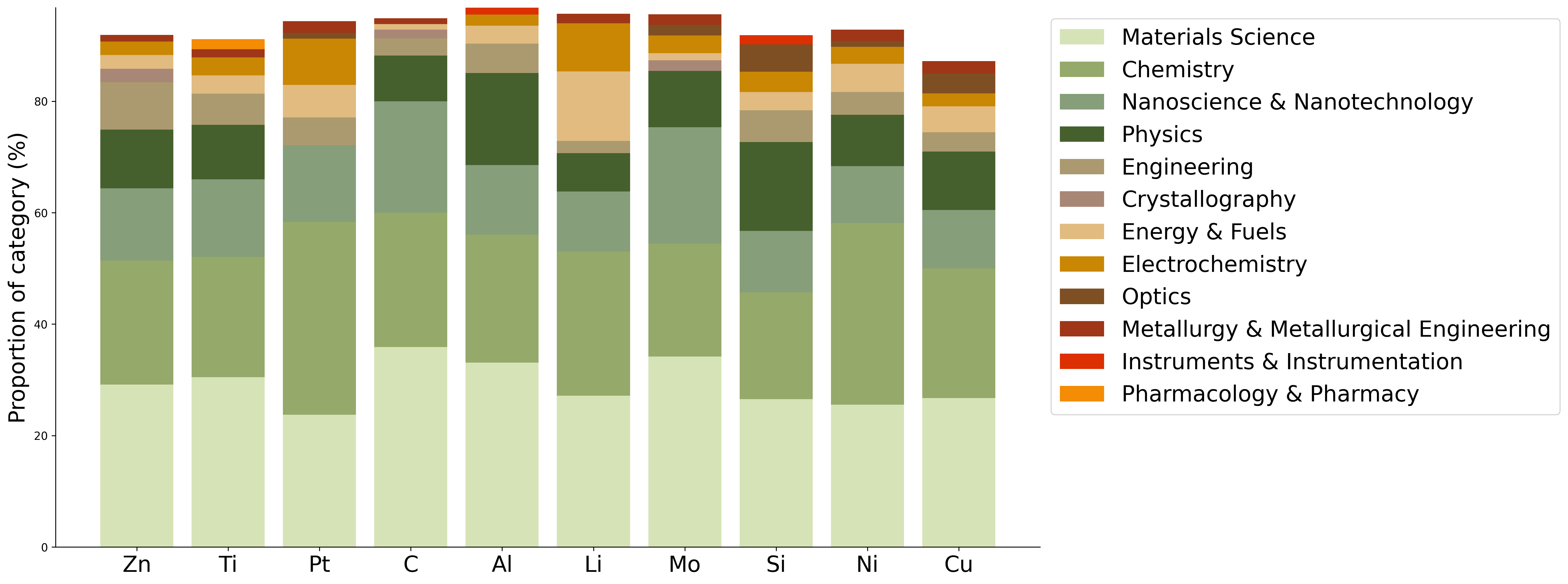}
  \caption{Proportion of publication scientific category of the popular ALD chemical elements}
\end{figure}
\begin{figure}[H]
\centering
\subfigure[]{
\label{Fig.sub.1}
\includegraphics[width=0.55\textwidth]{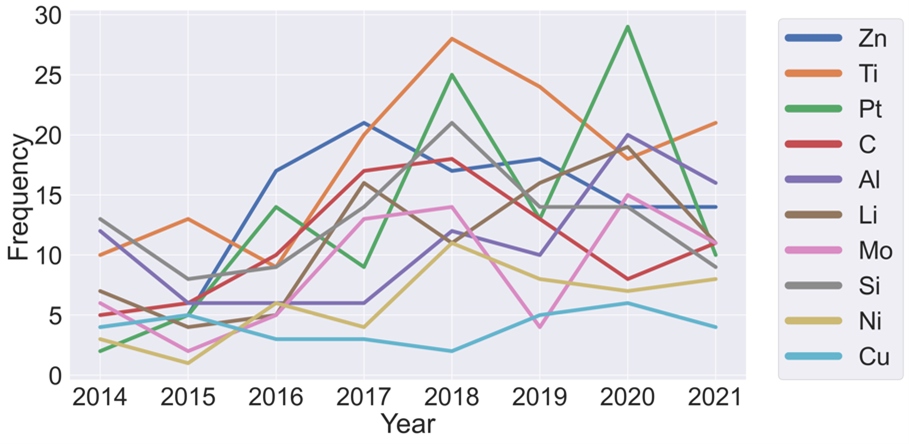}}
\subfigure[]{
\label{Fig.sub.2}
\includegraphics[width=0.25\textwidth]{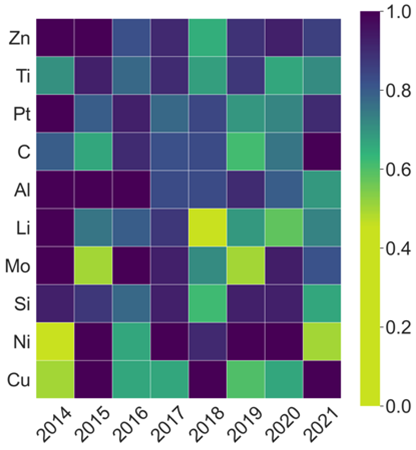}}
\caption{(a) Frequency and (b) sentiment heatmap of popular ALD chemical elements from 2014 to 2021 (sentiment scores vary from 0 to 1 and the darker the color, the more positive the sentiment)}
\label{Fig.main}
\end{figure}

\paragraph{Case study on Li}
We examined the scientific facts in extracted opinions of lithium, an element with marked changes in sentiment score. As we can see in Figure 4(b), lithium experienced a prominent negative sentiment in 2018 due to intrinsic properties, including thermal instability \cite{zhang2018significant,yu2018rational}, shuttling effects \cite{put2018chemistry}, and electrochemical instability \cite{xiao2018surface}. Fortunately, several improvement measures of these drawbacks were found in opinions of the following years. In 2019, Li \cite{li2019ultra} provided the possible solution for the \ce{LiMn1.5Ni0.4O4 (LMNO)} nanomaterial cycle time through selectively controlled methods, integrating ALD with annealing at the nano level and process controlling and optimization at the macro level. After that, Lou et al. \cite{lou2019ti} showed that the addition of Ti to Li-compounds can improve its cycling stability which can easily be achieved using ALD. These drawbacks and improvements were from diversified categories, such as energy fuels, nanoscience, and metallurgy. This case study shows that our system can further provide the reasons behind rising or falling sentiments. 

\subsection{CNN assisted investigation of potential nanomaterial textures for silicon solar cell}
The wide range of complex nanotextured surface morphologies made a comprehensive and systematic empirical investigation impossible. An unambiguous performance review for reported empirical results was necessary to determine the promising materials' nanotextures. For generating textual and graphical information for nanotextures, we implemented GraphMaster to help the data recovery from material nanotextures characterization results, which were usually present as figures, such as Raman spectroscopy spectra, and wavelength-dependent External Quantum Efficiency (EQE).

\paragraph{Case study of b-Si nanotextures for solar cells}
we used NLP, CV and ML to generate a highly refined database of reported b-Si solar cell results from the literature. The procedure was rapid and straightforward, and the generated database provided a comprehensive insight into b-Si nanotextures for solar cell applications. For example, the solar cell current improvement spectrum in Figure 5 was generated by 45 reported EQE spectrums from 17 independent research groups under certain criteria. It should be noted that the b-Si nanotextures used in the solar cells of these research works were highly distinct from each other. The figure shows that there is a solar cell current increase for the 300-600 nm wavelength range when replacing the conventional textures with the b-Si nanotextures. We can draw a solid conclusion that b-Si nanotextures are promising for solar cell applications as such textures will contribute to superior optical and electrical performance in the short-wavelength spectral region. 

 \begin{figure}[H]
  \centering
  \includegraphics[width=0.9\textwidth]{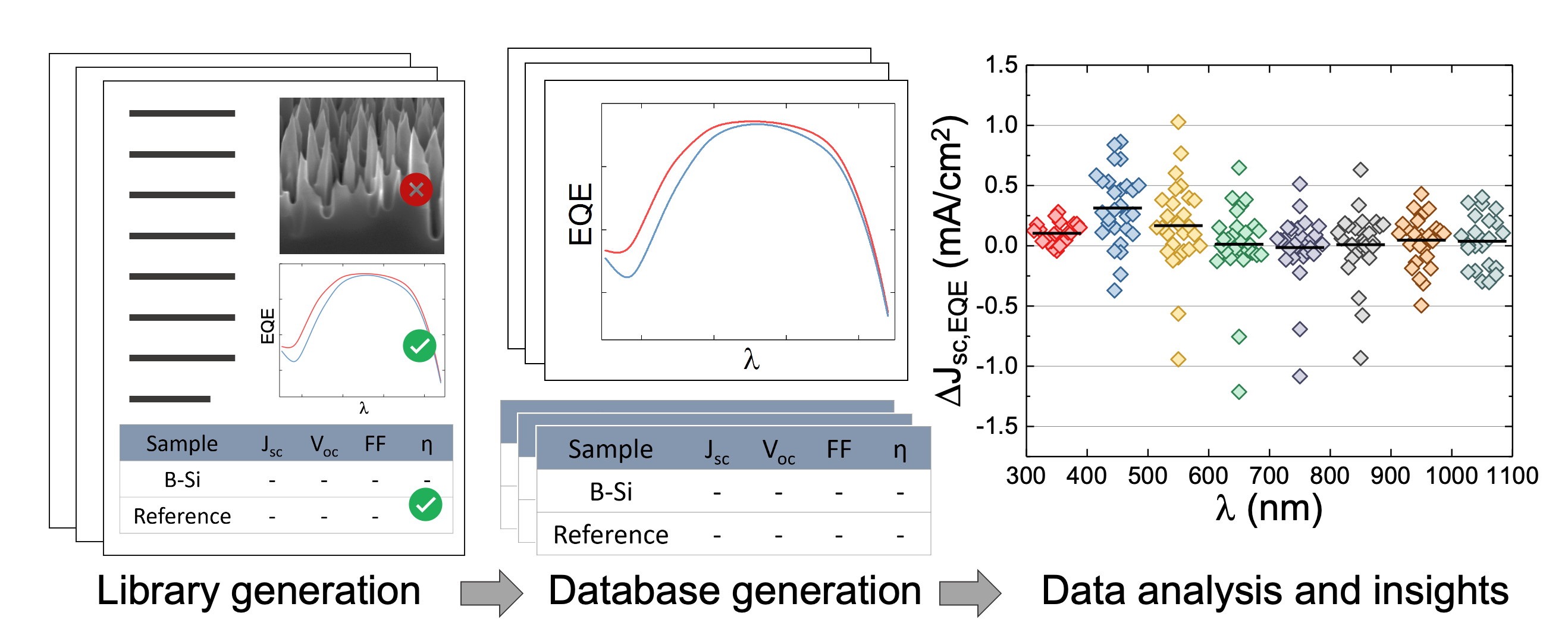}
  \caption{Schematic diagram of b-Si Nanotexture analysis pipeline. Reprinted with permission from \cite{wang2022artificial}. Copyright {2022} American Chemical Society.}
\end{figure}

\section{Conclusions and Future Work}
This research prepared a large nanomaterial literature collection with about 381,000 sentences and 34,000 graphs for model training. Two CNN models were developed using these data to extract scientific information from unstructured texts and graphs. Deep learning approaches helped us to predict new thermoelectrical materials and provided valuable solutions to the problems of existing nanomaterial synthesis. Furthermore, we also used the graph information combined with numerical simulation to compare the different nanotextures' performance. Our future work will address data recovery for more kinds of material characterization graphs. In summary, the results of both systems reported show that a CNN pre-classification process can improve scientific information extraction accuracy. The linkage between the text and graph information will provide a foundation to test and verify existing nanomaterial performance, which points to an interesting direction for future work. Our automated information extraction pipeline is expected to boost nanomaterial data mining research by replacing the heavy work of manual information extraction and we are also looking forward to building more downstream applications other than designing optimized material composition and synthesis methods.

\section{Acknowledgments}
This work was supported by funding from the Australian Centre for Advanced Photovoltaics(ACAP). The authors thank A/Prof. N.J. Ekins-Daukes for his GPU Resources from The Univeristy of New South Wales (UNSW). The authors also thank for the valuable suggestion from Laurence Berkeley National Lab (J.Dagdelen \& A.Dunn) and Carnegie Mellon University(H.Wen). T.Xie thanks the research scholarship from UNSW Materials \& Manufacturing Futures Insitute.

\bibliography{ref}
\clearpage
\appendix
\section{Textmaster experiment details} 
\begin{table}[H]
    \centering
    \begin{tabular}{ |c|c|c|c|c|c| } 
    \hline
    Task & Methods & Precision & Recall & F1-score & Accuracy \\
    \hline
    \multirow{9}{5em}{Opinion extraction} & TextBlob & 0.2460 & 0.3007 & 0.2706 & 0.6432 \\ 
    & Unsupervised-lexicon & 0.3525 & 0.3203 & 0.3356 & 0.7209 \\ 
    & Corpus comparison & 0.4451 & 0.5033 & 0.4724 & 0.7525 \\
    & CNN & 0.5614 & 0.6275 & 0.5926 & 0.8101\\
    & CNN-Attention & 0.5856 & 0.4248 & 0.4924 & 0.8072\\
    & BiLSTM & 0.5625 & 0.4706 & 0.5125 & 0.8029\\
    & CNN-LSTM & 0.5115 & 0.4379 & 0.4718 & 0.7842\\
    & BiGRU & 0.5319 & 0.4902 & 0.5102 & 0.7928\\
    & CNN-SMOTE & 0.6082 & 0.7712 & 0.6801 & 0.8403\\
    \hline
    \multirow{8}{5em}{Opinion classification} & TextBlob & 0.8211 & 0.8632 & 0.8416 & 0.7516 \\
    & Unsupervised-lexicon & 0.7257 & 0.7008 & 0.7130 & 0.5686 \\ 
    & Corpus comparison & 0.8977 & 0.6752 & 0.7707 & 0.6928 \\ 
    & CNN & 0.8947 & 0.8718 & 0.8831 & 0.8235\\
    & CNN-Attention & 0.9266 & 0.8632 & 0.8938 & 0.8431\\
    & BiLSTM & 0.8707 & 0.8632 & 0.8670 & 0.7974\\
    & CNN-LSTM & 0.9035 & 0.8803 & 0.8918 & 0.8366\\
    & BiGRU & 0.8807 & 0.8205 & 0.8496 & 0.7778\\
    \hline
    \end{tabular}
    \caption{Performance of different methods on test set of EoL dataset}
    \label{tab:my_label}
\end{table}
In opinion extraction task, precision here presents the fraction of correct predictions of opinion among the retrieved opinions, while recall is the fraction of correct predictions of opinion among all true opinions (because we focus more on opinions instead of non-opinions). In opinion classification task, precision here presents the fraction of correct predictions of opportunity among retrieved opportunities, while recall is the fraction of correct predictions of opportunity among all true opportunities (because we focus more on opportunities instead of challenges). F1-score is defined as the harmonic mean of precision and recall. The accuracy is the fraction of correct predictions among all predictions made. The SMOTE method was used to double the size of opinion category to further alleviate the data imbalance problem of opinion and non-opinion categories in the EoL dataset.

We implemented deep learning models with an NLP transfer learning framework called Kashgari \footnote{https://github.com/BrikerMan/Kashgari}. The word representations we used were publicly available 200-dimensional word embeddings \footnote{https://github.com/materialsintelligence/mat2vec}. As for the model's hyperparameters, we tried some combinations and selected the one with the highest accuracy on the development set: batch size = 64, and epochs = 30.

\section{Opinion mining results of perovskite and thermoelectric materials}
\begin{figure}[H]
\centering
\subfigure[]{
\label{Fig.sub.1}
\includegraphics[width=0.6\textwidth]{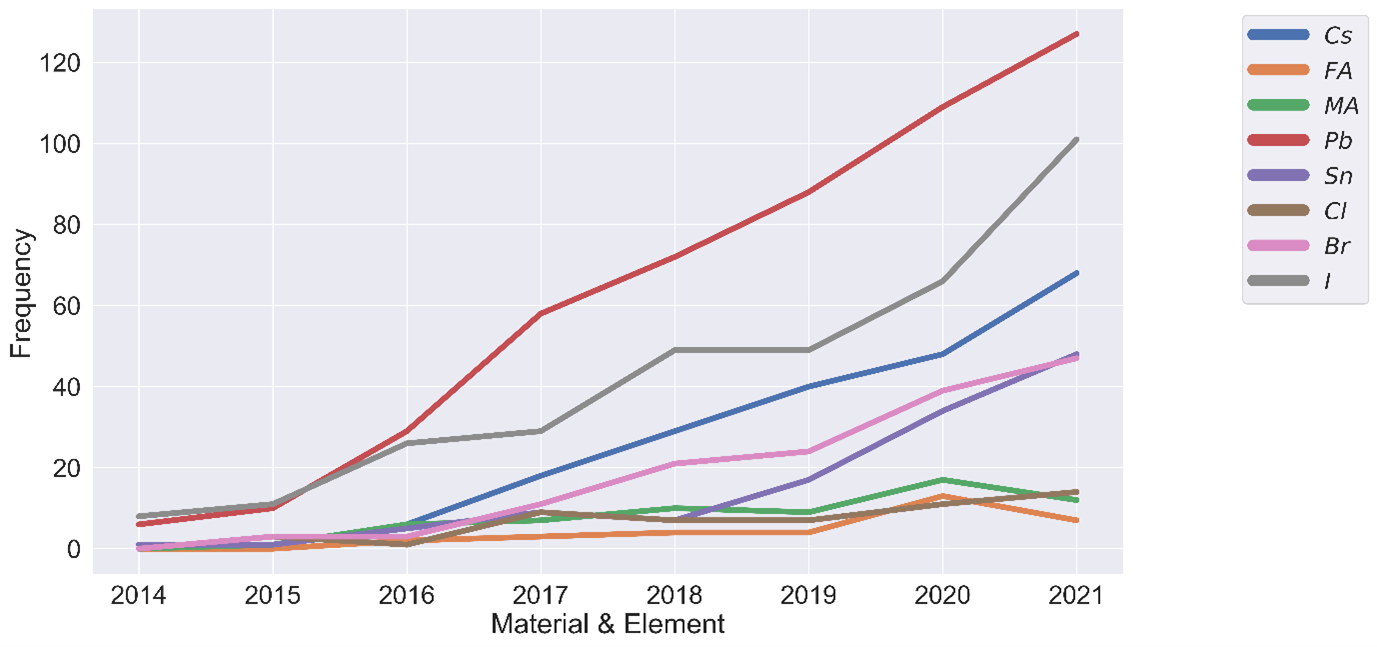}}
\subfigure[]{
\label{Fig.sub.2}
\includegraphics[width=0.3\textwidth]{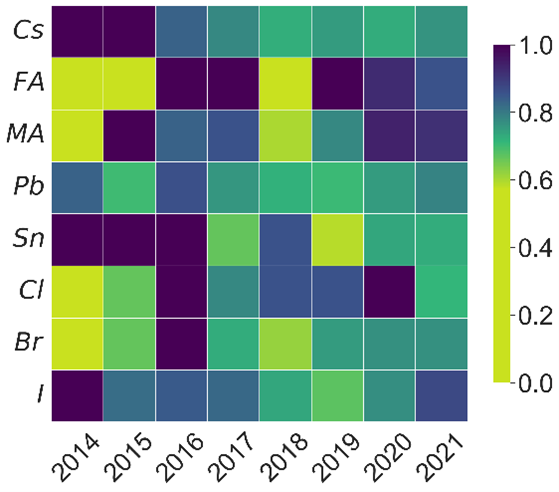}}
\caption{(a) Frequency and (b) sentiment heatmap of popular perovskite solar cell used chemical elements from 2014 to 2021}
\label{Fig.main}
\end{figure}

\begin{figure}[H]
\centering
\subfigure[]{
\label{Fig.sub.1}
\includegraphics[width=0.6\textwidth]{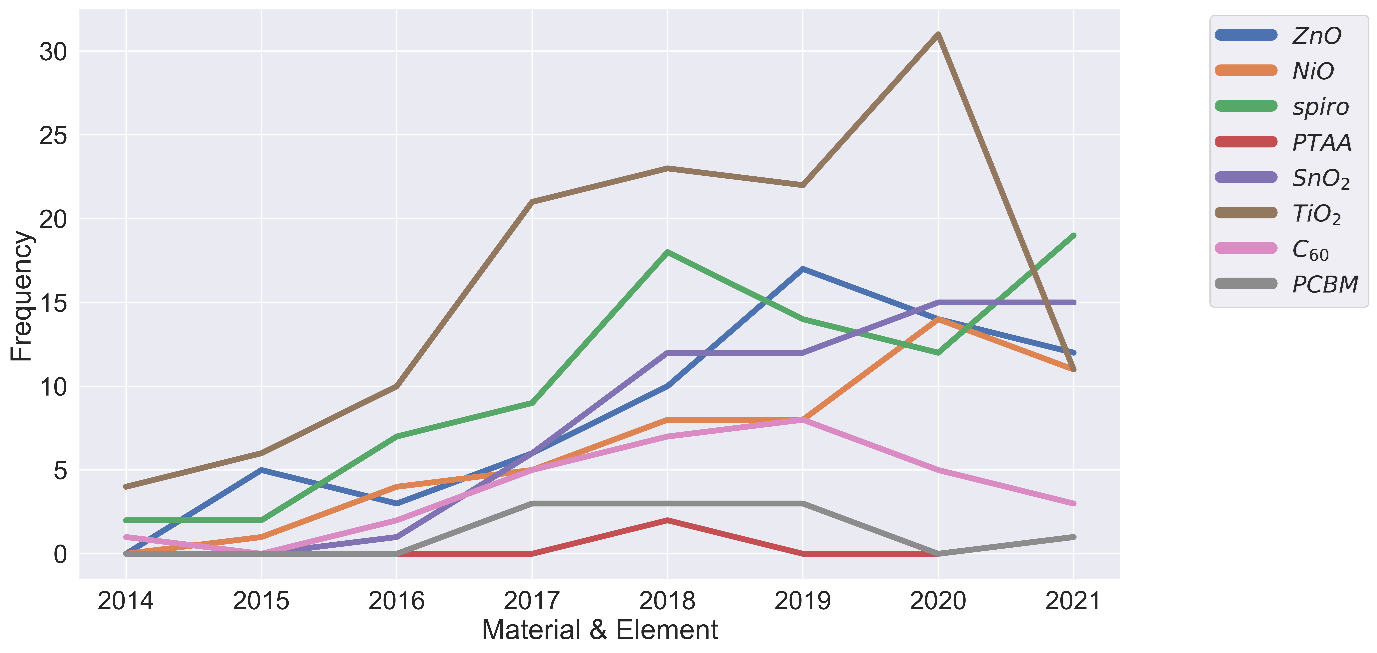}}
\subfigure[]{
\label{Fig.sub.2}
\includegraphics[width=0.3\textwidth]{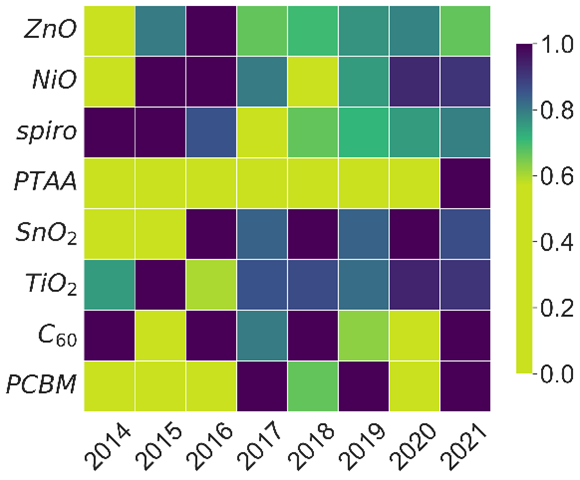}}
\caption{(a) Frequency and (b) sentiment heatmap of popular perovskite solar cell used materials from 2014 to 2021}
\label{Fig.main}
\end{figure}

\begin{figure}[H]
  \centering
  \includegraphics[width=0.7\textwidth]{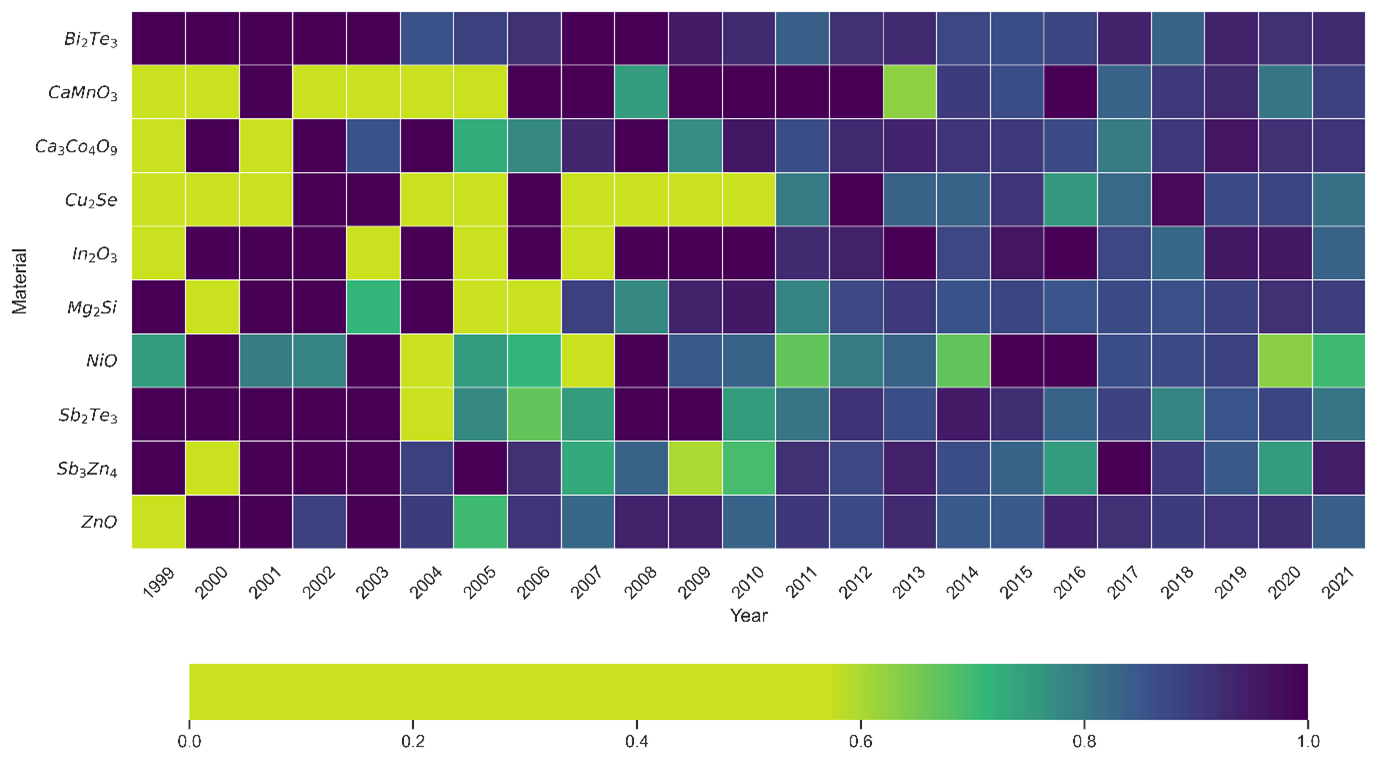}
  \caption{Sentiment heatmap of popular thermoelectric materials from 1999 to 2021}
\end{figure}

\section{Graphmaster Classifcation Example}
\begin{figure}[H]
  \centering
  \includegraphics[width=0.9\textwidth]{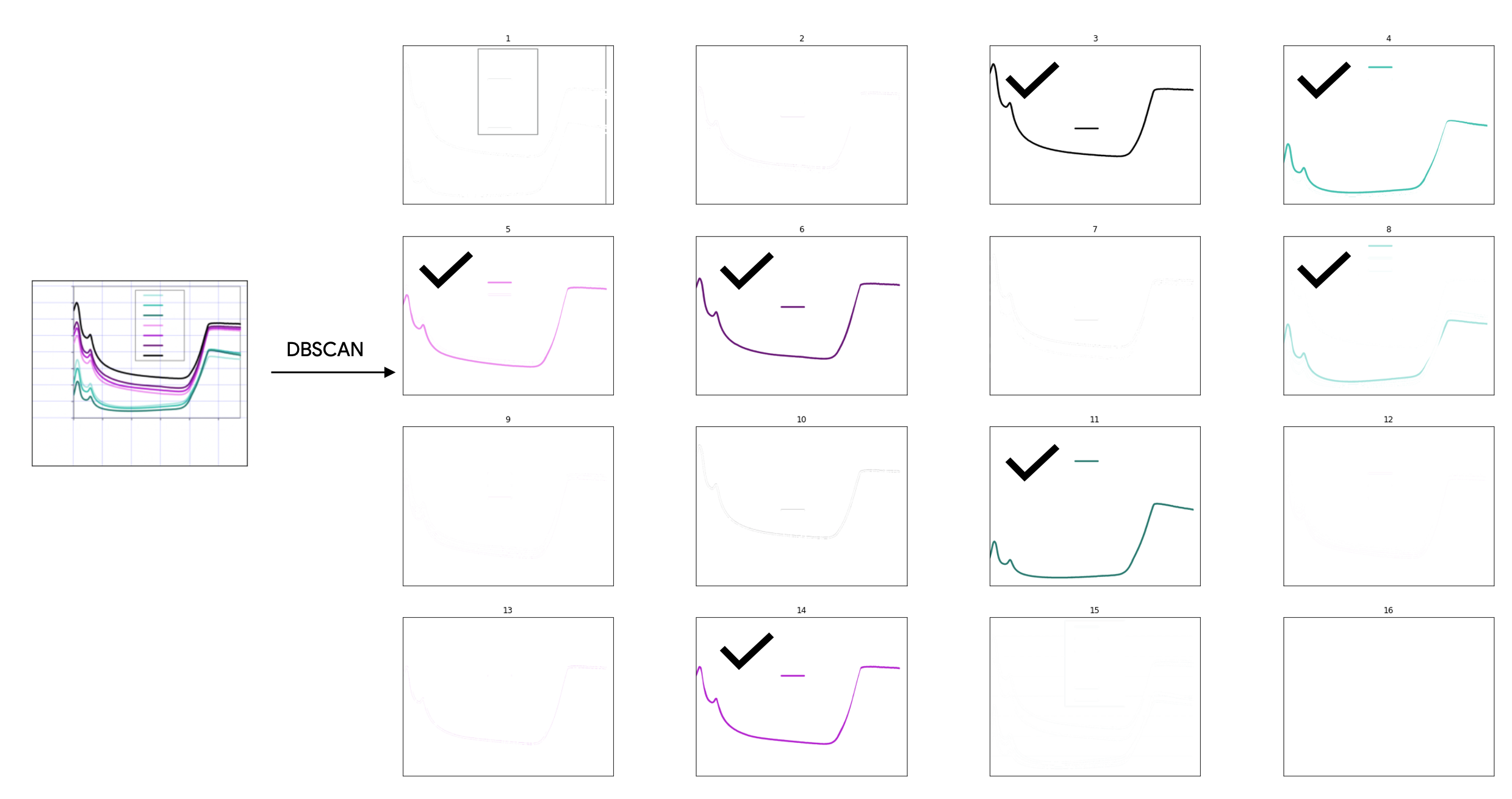}
  \caption{Process flow of Graphmaster Classification in example line charts. The figure with black ticks was the figure selected for the next step of data recovery. We excluded the figure with edge lines such as subfigure 1 by their locations.}
\end{figure}

\section{Data Recovery Results}
Here we showed the detailed result of the example in fig. 2. Since we clustered the figure based on different RGB values. Therefore, the figure contains not only the data line but also the legends and some edges of other lines. For example, in figure 2, the author used green colours from dark to light to annotate the R1-R3 and purple colours from dark to light to annotate M1 to M3. Therefore, The R3 line's edge colour was the same as the R1. To solve that problem, we scanned the figure from left to right and only selected the data with a minimum distance from the previous y-value when there were two more pixels in one column. 
\begin{figure}[H]
  \centering
  \includegraphics[width=1\textwidth]{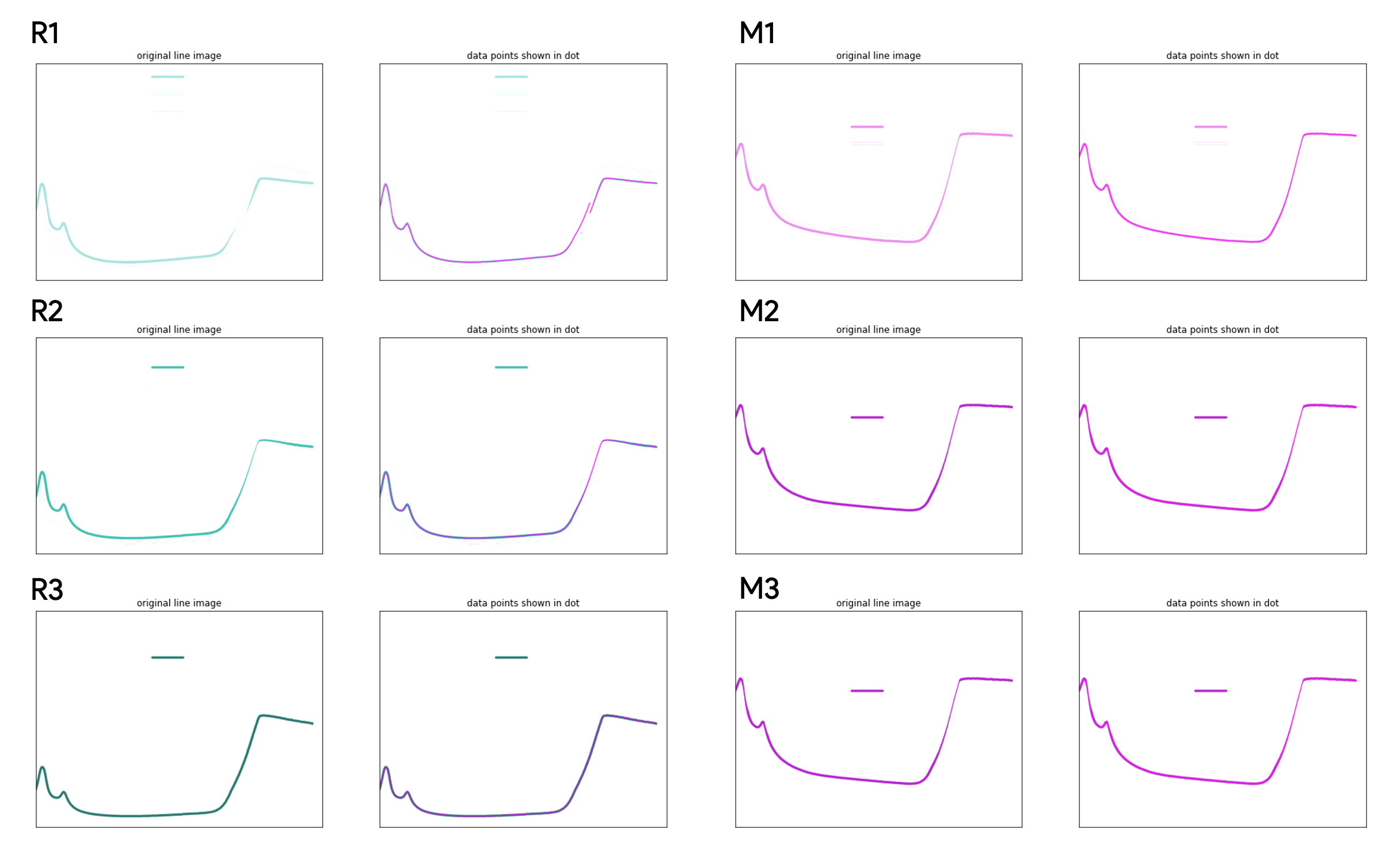}
  \caption{Details of data extraction in example line charts. We presented them with different legend labels. The left figure is the result after the colour cluster and the purple line in the right figure is plotted by the extracted data. }
\end{figure}

\section{Graphmaster Classification Experiment Result}
The classification step is trained on the  \href{https://researchweb.iiit.ac.in/~jobin.kv/projects/}{DocFigure dataset} with 21K various types of charts and images. The ResNet model is trained on 1 Nvidia Tesla V100-SXM2 GPU, with 32GB memory size.
\begin{table}[H]
\centering
\caption{Accuracies of different models on test dataset}
\begin{tabular}{X@{}llllll@{}} \toprule
        \multicolumn{6}{c}{\bfseries Accuracy}\\ \cmidrule(lr){3-6}
      Task & Methods & Line chart & Table image & Bar chart & Algorithm image \\ \midrule
\multirow{2}{*}{Graph clissification}   & ResNet      & 97.9\%    & 97.33\% & 99.19\% & 98.78\% \\
   & DenseNet      & 96.9\%    & 69.73\% & 71.88\% & 74.35\%  \\ 
\bottomrule
\end{tabular}
\end{table}
The text detection process was tested on 120 wavelength-dependent EQE graphs. 
\begin{table}[H]
\centering
\caption{Accuracies of text detection}
\begin{tabular}{X@{}llllll@{}} \toprule
        \multicolumn{6}{c}{\bfseries Accuracy}\\ \cmidrule(lr){2-6}
      Task & X-axis title & X-axis value & Y-axis title & Y-axis value & Legends \\ \midrule
\multirow{1}{*}{Text Detection}   & 98.3\%  & 96.8\% & 98.4\% & 93.4\% & 56.4\% \\
\bottomrule
\end{tabular}
\end{table}
We showed the mean square error of the example image presented in Fig.2 below:
\begin{table}[H]
\centering
\caption{Mean square errors of example figure}
\begin{tabular}{X@{}llllllll@{}} \toprule
        \multicolumn{8}{c}{\bfseries Mean Square Error}\\ \cmidrule(lr){2-8}
      Task & iso & M3 & M2 & M1 & R3 & R1 & R2 \\ \midrule
\multirow{1}{*}{Data Extaction}   & 5.6\%  & 2.5\% & 3.5\% & 3.5\% & 7.1\% & 5.8\% & 2.1\% \\
\bottomrule
\end{tabular}
\end{table}
\end{document}